# Nearest neighbor vector analysis of sdss dr5 galaxy distribution

**Yongfeng Wu[1*], Weike Xiao[2], Rongjun Mu[2] , David Batuski[3], Andre Khalil[4]**

[1]Department of Astronomy, University of Science and Technology of China, Hefei, P.R.China; [*]Corresponding Author: yongfeng.wu@maine.edu
[2](2nd Affiliation): Department of Astronautics Engineering, Harbin Institute of Technology, Harbin, P.R.China.
[3](3nd Affiliation): Department of Physics, University of Maine, Orono, ME, U.S.A.
[4](4nd Affiliation): Department of Mathematics, University of Maine, Orono, ME, U.S.A.

## ABSTRACT

**We present the Nearest Neighbor Distance (NND) analysis of SDSS DR5 galaxies. We give NND results for observed, mock and random sample, and discuss the differences. We find that the observed sample gives us a significantly stronger aggregation characteristic than the random samples. Moreover, we investigate the direction of NND and find that the direction has close relation with the size of the NND for the observed sample.**

## 1. INTRODUCTION

By the end of the 30s of last century, from the analysis of the position of galaxies on photographic film, Reference [1] found that the distribution of galaxies is not random and they aggregate obviously. Reference [2] statistically built simple galaxy clusters model with random distributions, but mismatched with the observation significantly. In the 50s, people [3, 4] observed thousands of clusters, many constituted by large numbers of galaxies. Even when the galaxies seemed isolated, they still have kind of correlation and can be described by correlation function, power spectrum, and other mathematical tools. The use of correlation functions to describe galaxy clustering has become widespread in recent years [5, 6]. The nearest neighbor [7, 8] distance is an especially powerful tool to describe small scales structures, because it depends on all the moments of the correlation function [9], thus it is extremely useful for revealing some aspects hidden in the correlation functions [7]. Even if it only provides information of the clusters pattern within a rather restricted range of scales [10], interesting results were obtained when this method was applied to galaxy data and also mock galaxy catalogs drawn from N-body simulations [11].

Compared to previous research, a significant difference of this paper is to develop the application of the Nearest Neighbor Vector (NNV) direction. Reference [12] mentioned that the angle of the two nearest neighbors of each galaxy can be used to discover the filaments. Here we regard the displacement between a galaxy and its nearest neighbor as a vector and discuss the direction distribution in the whole sphere. The motivation of this article is try to answer this question: is the universe clustered (by NND analysis)? If so, do galaxies have directional preference to select the nearest neighbor? By the size and direction analysis of the nearest neighbor, we could get more recognition about the hierarchical universe.

Our article is organized as follows. In section 2 we present the nearest neighbor statistical scheme. In section 3 we study the data from the SDSS DR5. We summarize the results in section 4 and have conclusions in section 5.

## 2. Nearest neighbor vector analysis

Reference [13] proposed the concept of the distance field. Suppose a given galaxy is in a three-dimensional coordinate system with Cartesian coordinates x, y, z. Let j be any other SDSS galaxy with Cartesian coordinates $x_j$ , $y_j$ , $z_j$ . For each galaxy the distance to its nearest neighboring object $r_i$ is computed as [14]

$$r_i = \min[\sqrt{(x-x_j)^2 + (y-y_j)^2 + (z-z_j)^2}\,] \qquad (1)$$

In this paper we simply use

$$\bar{r} = \frac{\sum r_i}{n} \quad (2)$$

to calculate the average NND (nearest neighbor distance) in each sample. $r_i$ is the NND for each point and n is the total number of particles. According to [15], the random sample should have a value $\bar{r}_E$ equal to $1/(2\sqrt{\rho})$, the ratio $R = \bar{r}/\bar{r}_E$ could be used to measure the deviation from the random sample. For a random sample R=1, for an extreme aggregation distribution (all points together), R=0.

We also consider the direction of each NND in the distance field; we call it NNV (nearest neighbor vectors). We first construct a sphere to include all samples and then split the whole sphere into 180 triangles and investigate the distribution of NNV passing through each triangle. By analyzing the anisotropy of the NNV, we can find the footprint of the filaments and compare different samples in this way. The detailed description is in section 4.

## 3. Data

The Sloan Digital Sky Survey (SDSS) is one of the most ambitious and influential surveys in the history of astronomy. It is a major multi-filter imaging and spectroscopic redshift survey using a dedicated 2.5-m wide-angle optical telescope at Apache Point Observatory in New Mexico, United States. We use the SDSS Data Release 5 as our galaxy sample, the detailed information (include the Redshift-distance formula, and a mock sample from Millennium Run Semianalytic Galaxy Catalogue [16] can be found from the paper of [17, 18]. About 35,700 galaxies have been used after applying volume-limiting selection (e.g., [19]), which will ensure that the selected galaxy sample is substantially complete to our absolute magnitude limit M=-19.9. See **Figure 1** for the geometry of the sample.

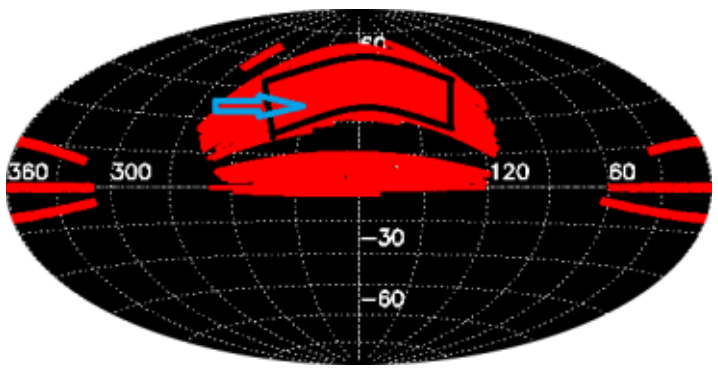

(a)

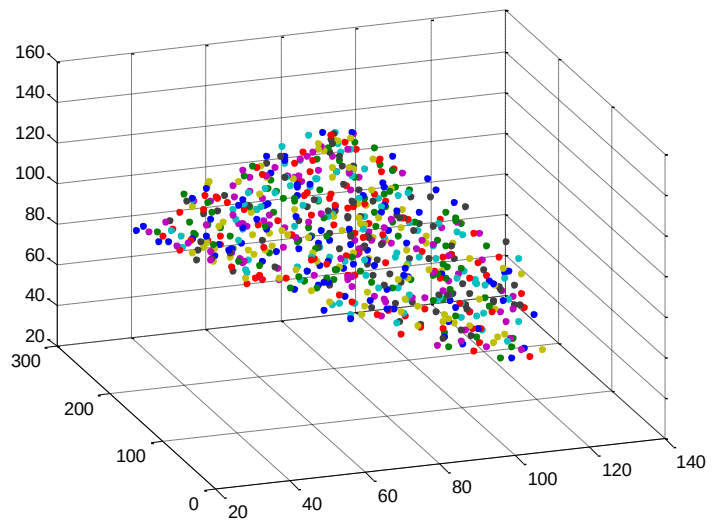

(b)

**Figure 1**. (a) SDSS sample geometry. The region inside the black "rectangle" of the figure is what we used. (b) 3D galaxy distribution, randomly (keep the original shape) selected hundreds galaxies from the observed sample.

## 4. Results

For the observed sample we get 1.95 Mpc for average nearest neighbor distance, for the mock sample, we get 2.3 Mpc, and for the random sample we get $r_E$ =3.5 $\pm$ 0.005 Mpc (11 random samples with different seeds). Then we have

$$R = \frac{\bar{r}}{\bar{r}_E} = \frac{1.95}{3.5} = 0.58 \qquad (3)$$

Clearly we can see observed sample has pronounced clustering on small scales compared with the random sample. Considering the extreme aggregation condition will have R=0 and random sample has an R=1, this observed sample is almost midway toward extreme aggregation. The clustering property of SDSS galaxies has been verified from various methods, such as two point correlation function [20, 21, 22]. The correlation length is about 5~7 Mpc [22] for Quasar and Luminous Red Galaxies (QSO-LRG). Our results of the average NND focus on more kinds of galaxies than QSO-LRG and support this clustering property on small scale from a new way. The mock sample has a R=0.66 in this measure value and is thus close to the observed sample.

Interestingly our analysis of the direction of the NND for each galaxy shows that the observed sample has an anisotropy property. To investigate the directional property of the NNV, we assumed all directions begin from a single point at origin, we split the whole surface of a sphere around the origin into 180 triangles (we could use the healpix method [23] to partition the surface into equal area "pixels", but the pixels are different triangular shape) displayed from the top to the bottom in the sequence, See following **Figure 2**.

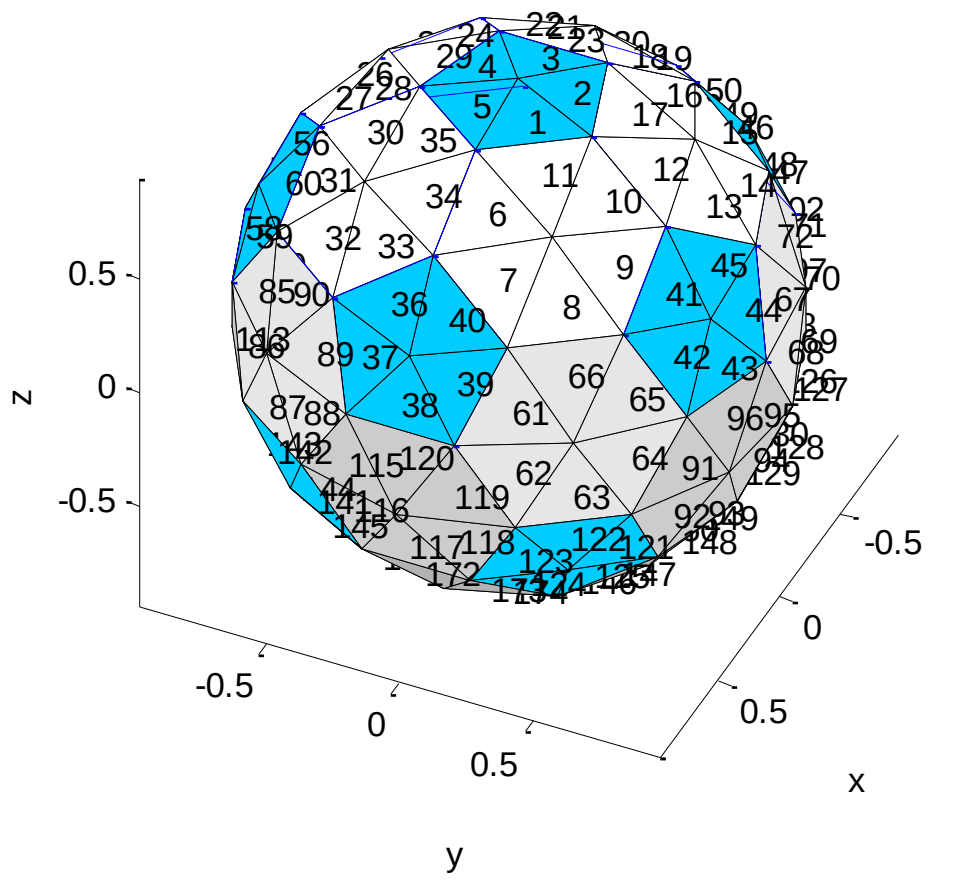

(a)

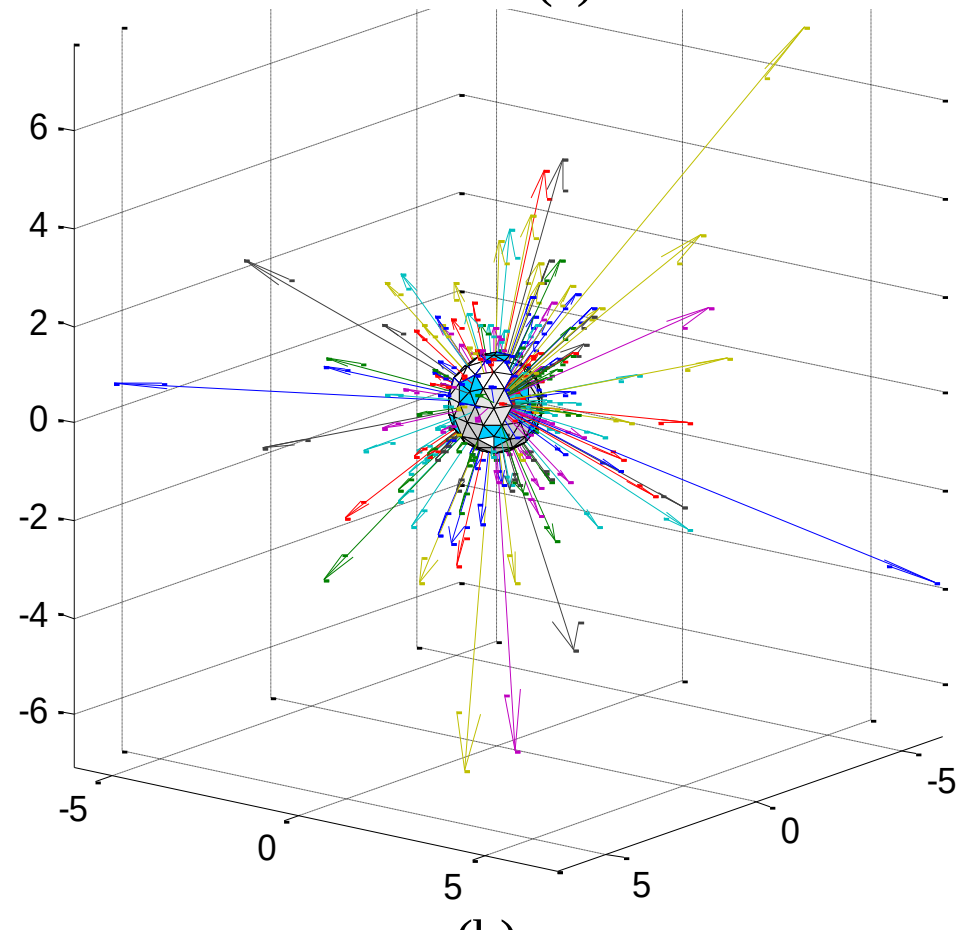

(b)

**Figure 2**: (a) The triangle surface of the sphere, the blue parts are pentagons and the white parts are hexagons. (b) An example of a NNV figure, each arrow represents the direction of the nearest neighbor of each galaxy and the length of the arrow represents the value of the NND, here we only plot 270 NNVs.

In **Figure 2a** we have 120 triangles belonging to the hexagons and 60 triangles belonging to the pentagons. They (such as triangle 2 and triangle 7) do not have the same area size as they belong to two different kinds of polygons, pentagon and hexagon. So we put a weight 1.26 on pentagon triangles to compensate for the smaller area comparable to the hexagon triangles, so the total number of NNV on all triangles will be around 8% larger than the total galaxies. If we plot all NNV of galaxies together (put the origin point at (0, 0, 0)), we can get a sketch of directions like **Figure 2b**.

We collect the NNV for all galaxies first, and then as we know the 3-D coordinates of the three vertexes of the each triangle and the direction of NNV, we could precisely calculate which nearest neighbor vector crosses and plot them with the sequence of 180 triangles and get the distribution in **Figure 3**.

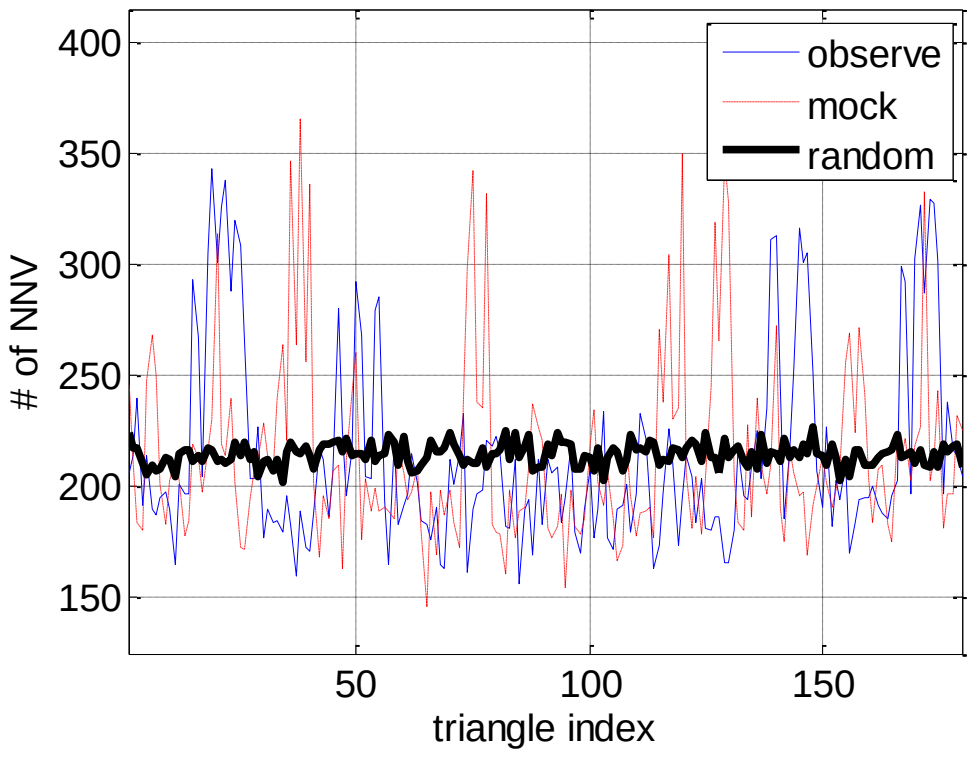

(a)

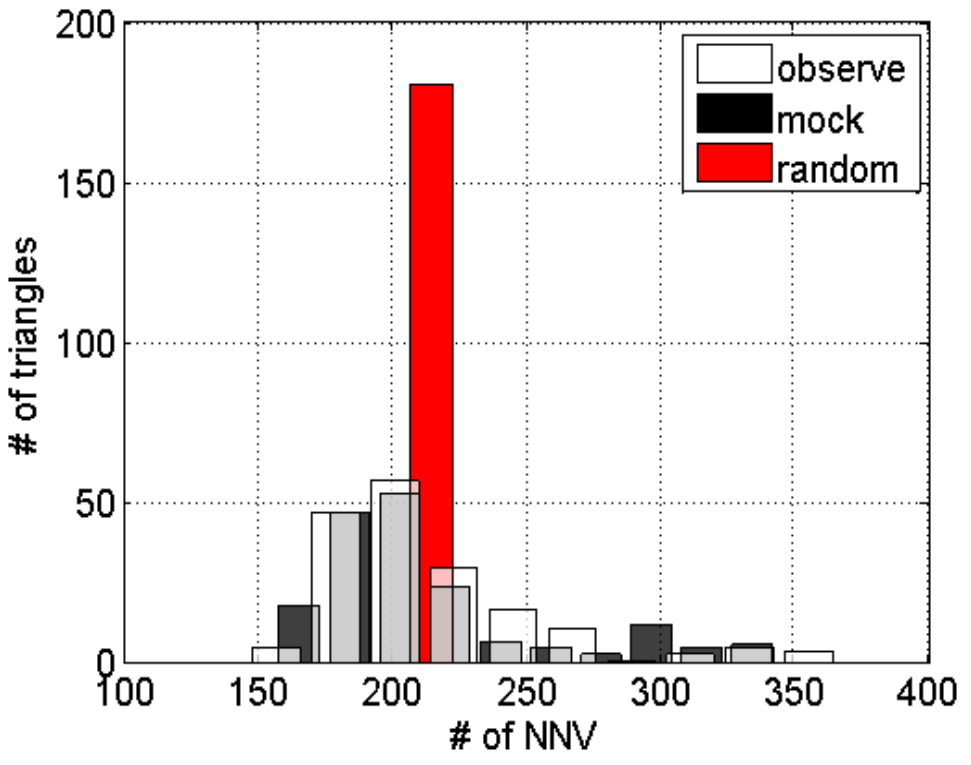

(b)

**Figure 3**: (a) NNV distributions on 180 angles for observe, mock, random samples. (b) Corresponding histogram for 3 samples. Overlap areas have gray color.

We also compute 11 random samples with different seeds to estimate the deviation. For all angles, mean value is 210 and we find that the average standard deviation ($\sigma$) is around 14 (maximum is 20) for all angles, in the following places all $\sigma$ are taken from here.

Here some peaks are separated only because the arrangement of 180 triangles is arbitrary, so even two adjacent triangles may have dozens of serial number difference. Observed sample and mock sample look very different at some specific triangles, but this is normal as the N-body simulation only simulates universe statistically, not exactly same with all details, such as the orientation of filaments. So we only focus on the global statistical properties from **Figure 3a**, not specific angles.

From **Figure 1**, our sample geometry looks like a distorted solid angle; how does this affect the NNV

analysis? **Figure 3** clearly tells us we do not need to worry about it as random samples have almost same distribution on all 180 triangles with the same geometry of observe and mock sample.

In **Figure 3** we clearly see that observed sample has a strong NNV distribution on some triangles, which are around triangle 20, 50 and corresponding opposite direction triangle 140 and 170 (for pairs of galaxies, two NNV directions are opposite). To investigate the relation between NNV and NND, we split galaxies into two groups, one has a smaller NND than average, and another has a larger NND than average. We plot them in **Figure 4** (two groups are normalized to have the same total number of NNVs).

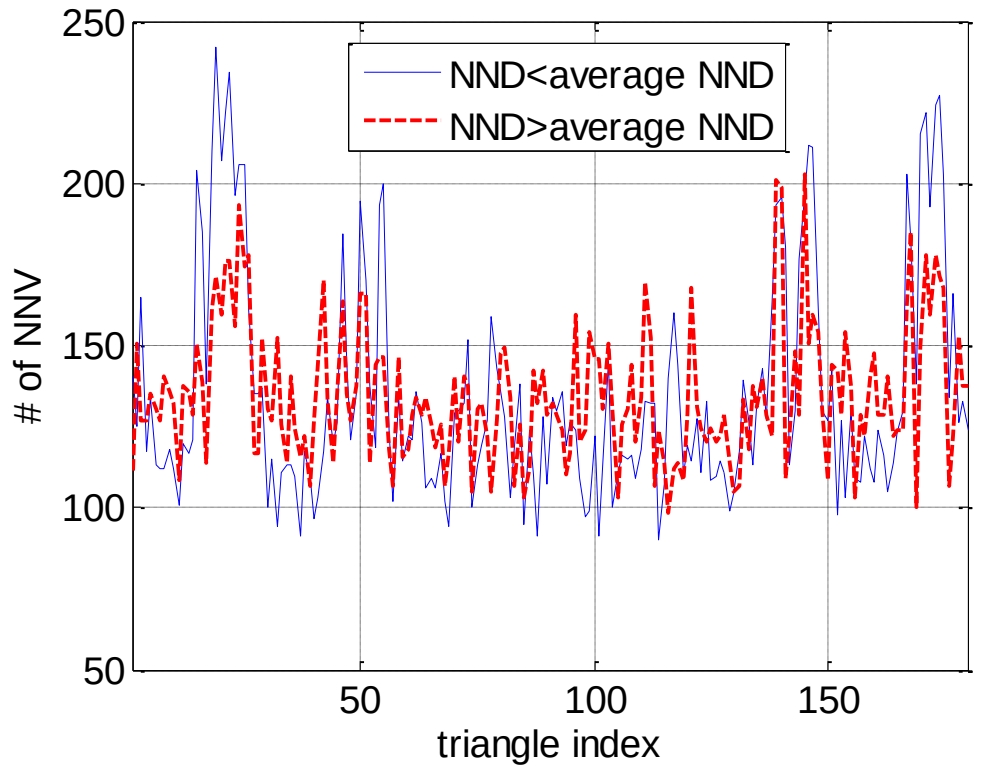

(a)

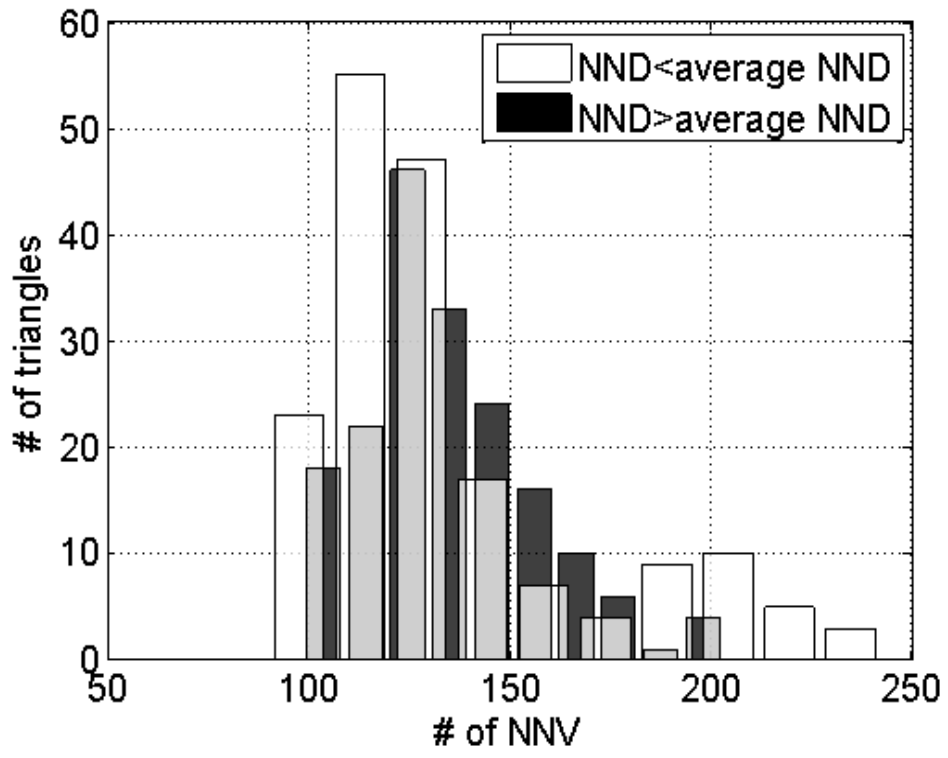

(b)

**Figure 4**: (a) NNV distribution for two kinds of galaxies: one kind has NND < 1.95 (average NND), another has NND >1.95. (b) Corresponding histogram for 2 kinds of sample.

We can see for smaller NND less than average, it displays a stronger anisotropy than galaxies, which have larger NND than average from **Figure 4a** & **4b.**

## 5. Conclusions

We have calculated the average NND of the SDSS galaxy sample and mock samples. We find the observed sample has a lightly smaller NND than mock sample, but much smaller than random sample. This result indicates that observed sample is more clustered in a special way. Moreover, we use a new method to investigate the direction distribution of NNV and find that the NNV of observed sample has a global anisotropy and is similar with mock sample, but clearly different from random sample on some angles. **Figure 3b** shows that the distribution of the random sample is like a delta function and this reflects the expected isotropic distribution. The result from the observed sample is more like a Poisson distribution and this leads us to think about the Gaussian fluctuations of cosmic microwave background (CMB). Both of them show the anisotropy resulting from the evolution of the universe, but with somewhat different statistical property. Maybe it is because our sample size is limited and needs further observations.

As both NNV and NND display significant difference between observed and random sample, this makes us think whether the NNV and NND are correlated. **Figure 4a** & **4b** shows that galaxies with smaller NND have stronger antistrophic NNV.

To better understand the physical sense of the results above, we shall check on the hypothesis about a global isotropic universe. There is a distinct hierarchy on a larger scale from a few hundred kpc to a few hundred Mpc [24]. Galaxies build up groups, clusters and superclusters, which in turn form a cellular structure of the Universe. We would expect that the observed sample has strong clustering property and a smaller NND than random sample; this is coincident with the NND results we find. However, the results of NNV reflect the anisotropy of a hierarchical universe in a very way more than the cluster property. Even in a much clustered point distribution, we still could get an isotropic NNV distribution, say, some symmetric spherical galaxy clusters, or some thin filaments (assume the thickness only includes one galaxy) uniformly distributed in all directions. So the NNV analysis provides a new way to distinguish how hierarchy is organized for the universe and we find galaxies do have a directional preference to select the nearest neighbor in universe.

.

## 6. ACKNOWLEDGEMENTS

The project is supported by key laboratory opening funding of the Harbin Institute of Technology (HIT.KLOF. 2012.077).